\documentclass[a4paper,11pt]{ivoa}
\newif\iftth
\iftth
\newcommand{\ivoaDocversion}{v2.0}
\newcommand{\ivoaDocdate}{2017-07-21}
\newcommand{\ivoaDocdatecode}{20170721}
\newcommand{\ivoaDoctype}{REC}
\newcommand{\ivoaDocname}{vtp}
\vcsrevision{a1e8d1c-dirty}
\vcsdate{2017-07-21 11:03:21 -0400}

\input tthntbib.sty

\begin{html}
<meta http-equiv="Content-type" content="text/html;charset=UTF-8"/>
\end{html}

\definecolor{ivoacolor}{rgb}{0.0,0.318,0.612}

\renewcommand{\author}[2][0]{\def\@tmp{#1}
  \if 0\@tmp
	{\begin{html}<li class="author">\end{html}#2\begin{html}</li>\end{html}}\else
	{\begin{html}<li class="author"><a href="#1">\end{html}#2\begin{html}</a></li>\end{html}}\fi}
\renewcommand{\previousversion}[2][0]{\def\@tmp{#1}
  \if 0\@tmp
	{\begin{html}<li class="previousversion">#2</li>\end{html}}\else
	{\begin{html}<li class="previousversion">
	  <a href="#1">#2</a></li>\end{html}}\fi}
\renewcommand{\ivoagroup}[1]
  {\begin{html}<dd id="ivoagroup">#1</dd>\end{html}}
\renewcommand{\editor}[2][0]{\def\@tmp{#1}
  \if 0\@tmp
        {\begin{html}<li class="editor">\end{html}#2\begin{html}</li>\end{html}}\else
        {\begin{html}<li class="editor"><a href="#1">\end{html}#2\begin{html}</a></li>\end{html}}\fi}

\newcommand{\includeMeta}{%
   \ivoaDocversion\ivoaDoctype\ivoaDocname\ivoaDocdate}

\def\SVN$#1: #2 ${%
	#2}

\newenvironment{abstract}{%
  \includeMeta
  \begin{html}
    </div> <!-- titlepage -->
    <div id="abstract"><h2>Abstract</h2>
  \end{html}
  }{%
    \ivoaDoctype
    \tableofcontents
  }

\newcommand{\lstloadlanguages}[1]{}
\newcommand{\lstset}[1]{}

\newenvironment{lstlisting}[1][plain]
  {\tthverbatim{lstlisting}}
  {}

\newcommand{\specialterm}[2]{%
  \begin{html}<span class="#1">\end{html}#2\begin{html}</span>\end{html}}
\newcommand{\xmlel}[1]{\specialterm{xmlel}{#1}}

\newcommand{\harvarditem}[4][0]{%
  
  \if 0#1 \item[#2 (#3)]
  \else \item[#1 (#3)]\fi}
\newcommand{\harvardurl}[1]{\url{#1}}

\def\AtBeginDocument#1{\relax}
\def\pgfsyspdfmark#1#2#3{\relax}

\newbox\voidb@x
\def\@m{\relax}

\begin{html}
  <div id="titlepage">
\end{html}

\fi

\usepackage[a4paper,left=1.2in,right=1.2in,top=1.2in,bottom=1.2in]{geometry}
\usepackage{listings}
\title{VOEvent Transport Protocol}
\ivoagroup{Data Access Layer / Time Domain Interest Group}
\author{Alasdair Allan}
\author{Robert B. Denny}
\author{John D. Swinbank}
\editor{John D. Swinbank}

\previousversion[http://www.ivoa.net/documents/Notes/VOEventTransport/]
                {http://www.ivoa.net/documents/Notes/VOEventTransport/}

\begin{document}

\begin{abstract}

The IVOA VOEvent Recommendation \citep{std:VOEVENT2} defines a means of
describing transient celestial events but, purposely, remains silent on the
topic of how those descriptions should be transmitted. This document
formalizes a TCP-based protocol for VOEvent transportation that has been in
use by members of the VOEvent community for several years and discusses the
topology of the event distribution network. It is intended to act as a
reference for the production of compliant protocol implementations.

\end{abstract}

\section*{Versioning}

The first version of the VOEvent Transport Protocol submitted for IVOA
approval is version 2.0. Earlier versions were described informally in an IVOA
Note \citep{Allan:2009}, which this document supersedes.

\section*{Acknowledgements}

Thanks to Robert W. White (ex LANL, now at NREL), Phillip Warner (ex NOAO),
Robert Seaman (NOAO). John D. Swinbank acknowledges support from the European
Research Council via Advanced Investigator Grant 247295.

\section{Introduction}
\label{sec:intro}

\begin{figure}
  \begin{center}
  \includegraphics{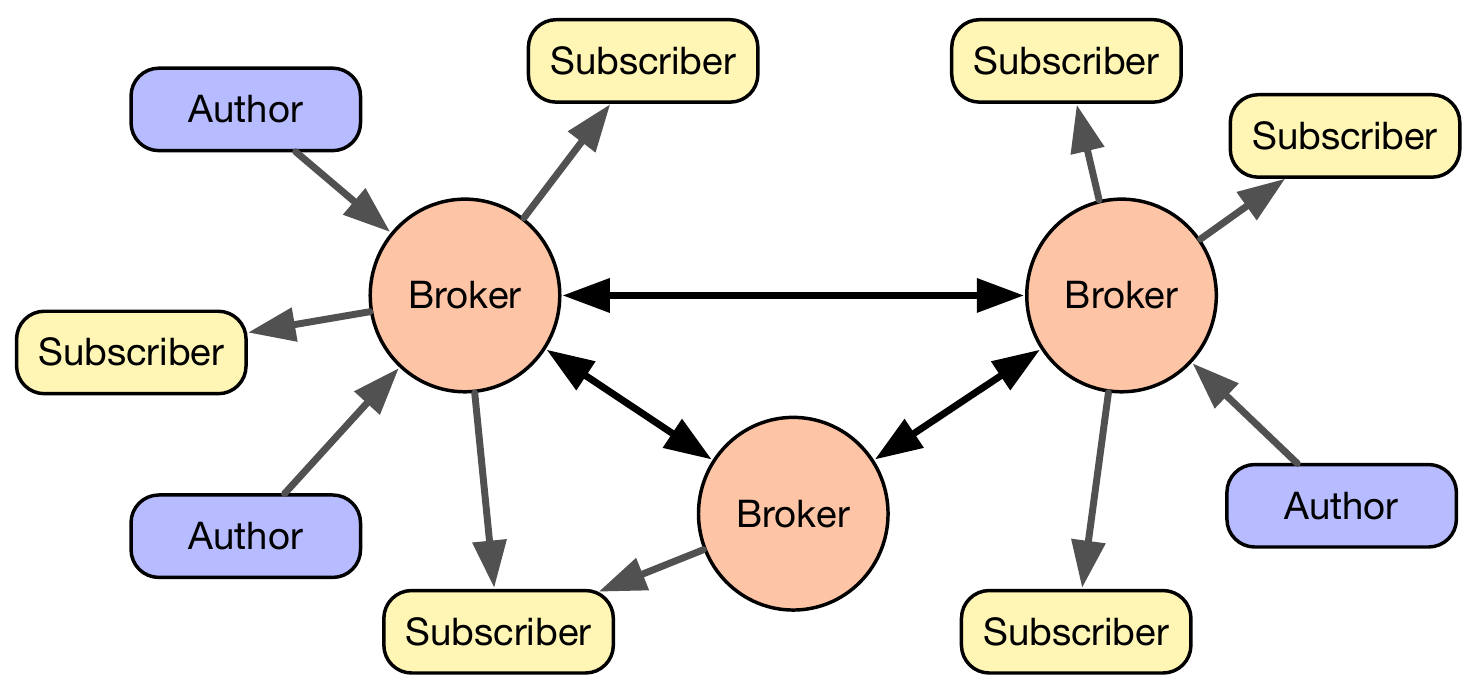}
  \end{center}

  \caption{VOEvent distribution system architecture showing the relationships
  between the various network roles.}

  \label{fig:network}
\end{figure}

The VOEvent standard \citep{std:VOEVENT2} defines a means of representing
transient celestial events with an implicit request for action on the part of
the recipient. The VOEvent standard is transport neutral: it does not take a
position on the mechanism by which the event should be transmitted from its
author to interested recipients. However, it encourages the construction of
``a robust general-purpose network of interoperating brokers'' for event
transmission.

To date, a number of different event distribution networks have been
prototyped and met with varying degrees of technical success and community
adoption. However, as the number of interested participants grows, and
next-generation large-scale survey instruments such as LSST\footnote{Large
Synoptic Survey Telescope; \url{http://www.lsst.org/}}, LIGO\footnote{Laser
Interferometric Gravitational-Wave Observatory; \url{http://www.ligo.org/}},
LOFAR\footnote{Low Frequency Array; \url{http://www.lofar.org/}} and
SKA\footnote{Square Kilometre Array; \url{http://www.ska-telescope.org/}},
which promise event rates ranging up to the millions per day, are
developed and begin to become available, it is clear that a standard,
interoperable mechanism for event communication is required. It is such a
mechanism that this document describes.

\begin{figure}
  \begin{center}
  \includegraphics[width=1.0\textwidth]{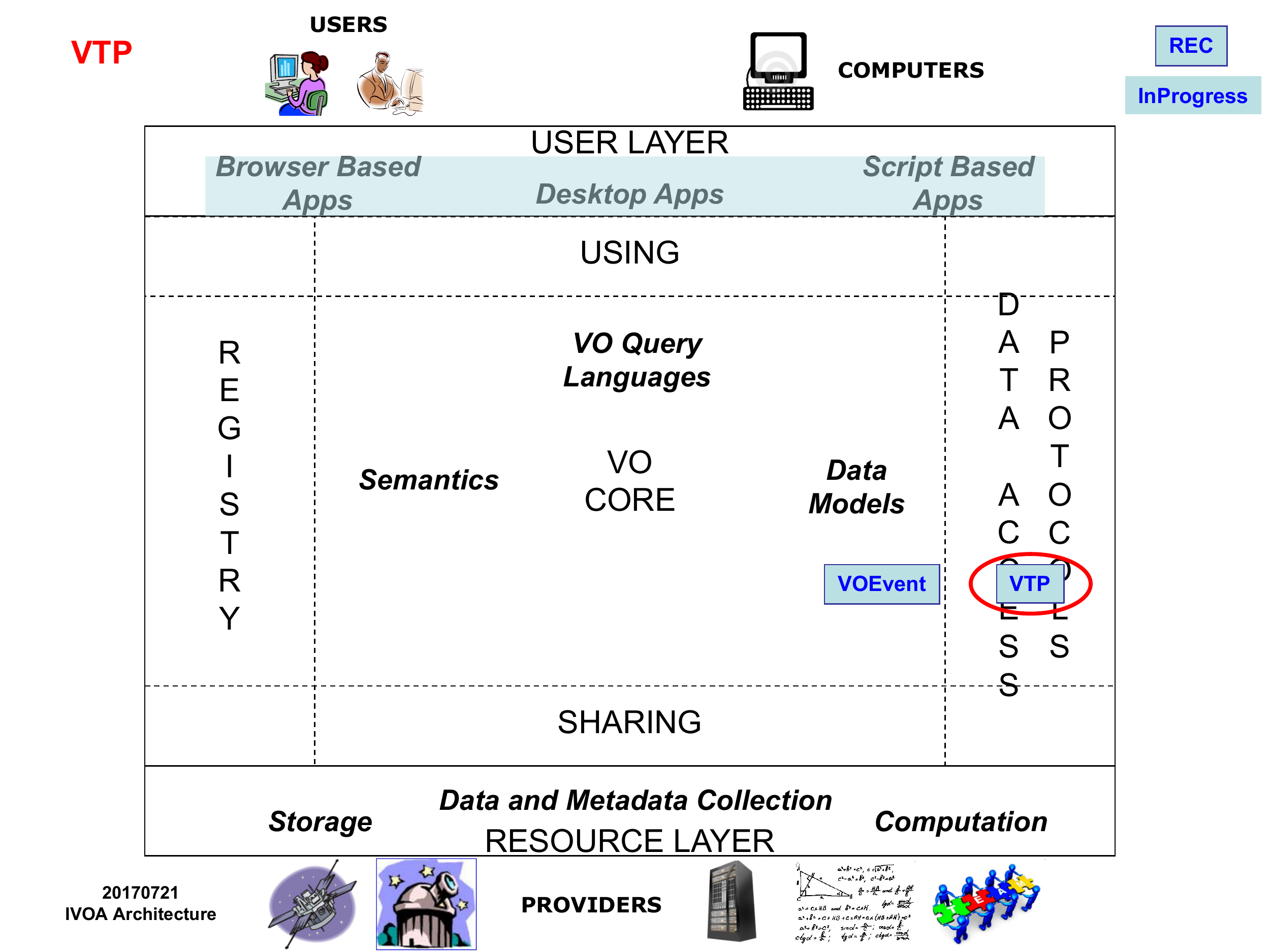}
  \end{center}

  \caption{Diagram based on \citet{note:VOARCH} showing the VOEvent Transport
  Protocol (VTP) in the context of the wider IVOA architecture.}

  \label{fig:ivoa_arch}
\end{figure}

The purpose of the protocol described herein is to transport a VOEvent
document from its sender to one or more interested recipients. To achieve
this, we envision three distinct network roles: authors, which create events;
brokers, which receive events from authors and distribute them, and
subscribers, which receive and (if appropriate) act upon the events. Refer to
Figure \ref{fig:network} for an illustration, while Figure \ref{fig:ivoa_arch}
shows how this integrates with the wider IVOA architecture. Note that a single
entity may perform more than one role within the network: for example,
creating events and distributing its own creations (combining the author and
broker roles) or receiving events from a broker and redistributing them to a
list of subscribers (combining the subscriber and broker roles).

Building upon this architecture, a strongly-connected set of brokers which
subscribe to each other's event streams and redistribute to their subscribers
(the ``VOEventNet backbone'') provides a fault-tolerant system which is
resilient against the failure of one or more network entities. Such a backbone
system is already under construction by members of the VOEvent community.

The protocol described herein is intentionally as simple as possible while
still accomplishing the required task. More complex protocols will be required
for addressing advanced use-cases, handling extremely large event or
subscriber numbers\footnote{Refer to \citet{Swinbank:2014} for a discussion of
scalability}, or providing value added services\footnote{For example, Svom,
the Space-based multi-band astronomical Variable Objects Monitor,
\url{http://www.svom.fr/}, is developing an approach based on XMPP
\citep{SaintAndre:2011}.}. These fall outside the scope of the current
document.

Although this document refers specifically to VOEvents, the protocol places
only quite minimal requirements on the payload. We expect that a future
evolution of this protocol would provide a convenient means of delivering
diverse message types, perhaps including portfolios or containers of VOEvents,
or even non-IVOA standard messages.

\section{Terminology}

Throughout this document, we adopt the terminology of RFC 2119
\citep{Bradner:1997}. In particular:

\begin{itemize}
    \item{The word ``must'' indicates an absolute requirement of the
    specification;}

    \item{The word ``should'' indicates behaviour that is normally included in
    implementations of the specification, but there may exist valid reasons
    for excluding it in particular circumstances;}

    \item{The word ``may'' indicates purely optional behaviour which is
    permitted according to this specification.}
\end{itemize}

\section{Common characteristics}

\subsection{Design goals}
\label{sec:common:design}

The VOEvent Transport Protocol, hereafter VTP, provides a simple means of
transporting VOEvent documents from authors through brokers to subscribers.

VTP transmits no more than one VOEvent in each transaction. If multiple
documents are to be transmitted, multiple transactions must take place.

VTP delivers VOEvents to eligible subscribers which exist on the network at
the time of transmission. It does not buffer events for later transmission.
Subscribers who wish to retrieve historical events should consult an event
repository.

VTP is non-transformational on VOEvents being transmitted: the document
delivered to a subscriber should be bit-for-bit identical to that provided by
an author. If an intermediary wishes to modify or annotate the VOEvent, they
should not edit the document in transport, but rather generate a new document
to supplement or replace it.

VTP does not provide a transport-level means of annotating or otherwise
embellishing VOEvent documents, or of providing stream-level metadata.

VTP values simplicity of design and operation to lower the barrier to entry. It
is not intended to meet every use case. As per Section \ref{sec:intro}, it is
anticipated that some VOEvent-based services will require more complex
protocols.

VTP is independent of implementation: conforming network entities should be
able to interoperate seamlessly, even when derived from different codebases.

\subsection{Network layer}

VTP operates over TCP \citep{Cerf:1974} connections, and relies on TCP's
guaranteed error-free in-order delivery of data: no checksum or digest data is
included. All documents are sent over the TCP connection preceded by a 4-byte
network-ordered\footnote{As defined by \citet{Reynolds:1994}; also called
``big-endian'' ordering.} count, followed immediately by the payload data. The
4-byte count is interpreted as a 32-bit integer equal to the number of payload
bytes following the count bytes. The payload is considered an opaque
collection of bytes at this level\footnote{As a result, the format of the
document being transmitted is opaque to the transport layer. Therefore both
ASCII and UTF-8 are equally supported}.

\subsection{Message format}
\label{sec:common:format}

Throughout this document, the term ``message'' refers to a complete VTP
message, including both the initial byte count and the message payload.  The
payload is an XML document. It must consist of an XML declaration followed by,
in order, optional XML comments, a single \xmlel{<VOEvent~/>} or
\xmlel{<Transport~/>}\footnote{\xmlel{<Transport~/>} elements are used by
the VTP system itself and are invisible to end-users: see Section
\ref{sec:transport} for details.} element, and more optional XML comments. It
must validate against either the VOEvent XML
schema\footnote{\url{http://www.ivoa.net/xml/VOEvent/VOEvent-v2.0.xsd}} or the
Transport XML schema (Appendix \ref{sec:transportschema}). Messages may be
conveniently referred to by their payload type (viz. ``VOEvent message'',
``Transport message'').

\subsection{Broker behaviour}
\label{sec:common:broker}

Although the simplest broker implementation may simply forward all unique
events it receives, either directly from authors or from other brokers, to all
of its subscribers, this behaviour is not required. Instead, the broker may
provide ``added-value'' services which limit how messages are redistributed.
For example, a broker may make arrangements with some or all of its
subscribers to filter the events it receives, and forwarding only those events
that fulfil some predefined criteria. Similarly, brokers may limit access to
some clients based on various criteria (\S\ref{sec:limit}).

VTP does not provide in-band notification of these per-broker details. For
example, the protocol does not make an author submitting to a filtering broker
aware that their event might not be sent to all of the broker's subscribers,
and, similarly, it does not make a subscriber of a filtering broker aware that
they might not receive a complete set of events. It is the responsibility of
authors and subscribers to ensure that the brokers they use provide the
services they require. Brokers should clearly advertise any added-value
behaviour they provide, for example on a website or through the IVOA registry
\citep{note:VOARCH}.

\section{Network nodes}
\label{sec:node}

The VOEvent network consists of three types of nodes (refer to Fig.
\ref{fig:network}):

\begin{itemize}
    \item{Author}
    \item{Broker}
    \item{Subscriber}
\end{itemize}

As described in \citet{std:VOEVENT2}, it is expected that authors and brokers
will be registered with the IVOA registry\footnote{This is dependent on the
VOEvent Registry Extensions \citep{std:VOEventRegExt}, which are not fully
standardized or widely deployed at time of writing. Unregistered services may
therefore be deployed until such time as the relevant registry support becomes
available.}. It is not necessary for subscribers to register.

The flow of messages is over three types of connections:

\begin{itemize}
    \item{Author to Broker}
    \item{Broker to Subscriber}
    \item{Broker to Broker}
\end{itemize}

Each type of connection is discussed qualitatively below.

\subsection{Author to Broker}

When an author wants to submit a VOEvent document to the network, it
constructs a message encapsulating that document, opens a TCP connection to a
broker, sends the message, waits for a response from the broker, and then
closes the TCP connection. The response from the broker is a message
containing a Transport document.

\subsection{Broker to Subscriber}

When a subscriber wants to receive VOEvent traffic, it opens a TCP connection
to a broker. This connection is kept open continuously. When the broker
receives a VOEvent message, it relays a copy of that message to each connected
subscriber\footnote{If filtering as described in \S\ref{sec:common:broker} is
being carried out, the message may be sent only to a subset of the
subscribers.}.  Thus, a subscriber must continuously listen on the TCP
connection and be prepared to receive new messages at any time, even when it
is busy processing a previously received message. When a subscriber receives a
VOEvent message from its broker, it must respond with an appropriate Transport
message.

\subsection{Broker to Broker}
\label{sec:node:brokertobroker}

Traffic between brokers uses the preceding methods. Each broker takes the role
subscriber as far as every other broker is concerned. A broker that wishes to
receive a feed from another broker should connect to that broker's subscriber
port. No special protocol features are needed.

\section{Connection Maintenance}
\label{sec:maintenance}

All connections over which a broker sends VOEvent messages are kept open
continuously. However, basic TCP does not provide any dead-peer
indication\footnote{ TCP does support a ``keep-alive'' service, but it is not
universally available \citep{Braden:1989}.}. Further, network infrastructure
devices might sever a TCP connection after some period of inactivity. This
gives rise to the need for keep-alive messages. After no more than 90 seconds
of inactivity on any given connection, the broker must send a Transport
\xmlel{iamalive} message, to which the subscriber must reply with a copy of
that message plus some optional identification information. The message format
is described in Sections \ref{sec:transport:iamalive} and
\ref{sec:transport:iamaliveresponse}.

At both ends of the continuous connection, the node either expects to receive
an \xmlel{iamalive} message or expects to receive the response to its
\xmlel{iamalive} message. If not seen, the node should assume that the
connection has been lost or the peer is dead. At this point, the node that was
responsible for opening the connection may attempt to re-initiate it. The use
of geometric back-off algorithm may help alleviate network load.

\section{Transport messages}
\label{sec:transport}

Transport messages are VTP messages \S\ref{sec:common:format} containing a
\xmlel{<Transport~/>} element. There are four classes of
\xmlel{<Transport~/>} element, distinguished by their \xmlel{role}
attribute:

\begin{itemize}
\item{\xmlel{iamalive} (Connection maintenance);}
\item{\xmlel{authenticate} (Authentication request/response);}
\item{\xmlel{ack} (VOEvent successful receipt acknowledgement);}
\item{\xmlel{nak} (VOEvent unsuccessful receipt acknowledgement).}
\end{itemize}

All Transport messages have the same general syntax, and are defined by the
Transport schema (Appendix \ref{sec:transportschema}). The connection
maintenance and receipt acknowledgement message types are described in detail
in this section; the authentication message type has a special role which is
described in Section \ref{sec:limit:crypto}.

\subsection{\xmlel{iamalive} message}
\label{sec:transport:iamalive}

The \xmlel{iamalive} message is indicated by a role equal to \xmlel{iamalive}.
The \xmlel{<Origin~/>} element contains the IVOID\footnote{International
Virtual Observatory identifier; \citet{std:VOID2}.} of the broker which is
managing the connection. The \xmlel{<TimeStamp~/>} element contains the date
and time at which the message was generated formatted as per \S3.3.7 of
\citet{Peterson:2012}. This time should be provided in UTC, and may include a
“Z” timezone indicator.

\begin{lstlisting}[language=xml,caption=Sample \xmlel{iamalive} message.,
                   label=lst:iamalive]
<?xml version="1.0" encoding="UTF-8"?>

<trn:Transport role="iamalive" version="1.0"
 xmlns:trn="http://telescope-networks.org/schema/Transport/v1.1"
 xmlns:xsi="http://www.w3.org/2001/XMLSchema-instance"
 xsi:schemaLocation="http://ivoa.net/xml/Transport/v1.1
                     http://ivoa.net/xml/Transport-v1.1.xsd">
    <Origin>ivo://invalid.broker/example#</Origin>
    <TimeStamp>2001-01-01T00:00:00Z</TimeStamp>
</trn:Transport>
\end{lstlisting}

\subsection{\xmlel{iamalive} response}
\label{sec:transport:iamaliveresponse}

The \xmlel{iamalive} response is an extension of the initial \xmlel{iamalive}
message. It also has a role of \xmlel{iamalive}. The \xmlel{<Origin ~/>}
element is preserved unchanged from the \xmlel{iamalive} being responded to
(that is, it contains the IVOID of the broker). It may include an additional
\xmlel{<Response~/>} element containing a URI identifying the
subscriber\footnote{If the subscriber is registered with the IVOA registry,
this may be an IVOID, but registration is not required.}. It may also include
a \xmlel{<Meta~/>} element with \xmlel{<Param~/>} sub-elements which give
additional information about the subscriber or any other relevant information.
\xmlel{<Param~/>} elements have no content and must contain name and value
attributes. The names and values may be any string. The \xmlel{<TimeStamp~/>}
element contains the date and time at which the response was generated,
formatted as per \S3.3.7 of \citet{Peterson:2012}. This time should be
provided in UTC, and may include a “Z” timezone indicator.

\begin{lstlisting}[language=XML,caption=Sample \xmlel{iamalive} response.,
                   label=lst:iamaliveresponse]
<?xml version='1.0' encoding='UTF-8'?>

<trn:Transport role="iamalive" version="1.0"
 xmlns:trn="http://telescope-networks.org/schema/Transport/v1.1"
 xmlns:xsi="http://www.w3.org/2001/XMLSchema-instance"
 xsi:schemaLocation="http://ivoa.net/xml/Transport/v1.1
                     http://ivoa.net/xml/Transport-v1.1.xsd">
    <Origin>ivo://invalid.broker/example#</Origin>
    <Response>ivo://invalid.subscriber/example#</Response>
    <TimeStamp>2001-01-01T00:00:00Z</TimeStamp>
    <Meta>
        <Param name="IPAddr" value="10.0.0.0" />
        <Param name="Contact" value="name@subscriber.invalid" />
    </Meta>
</trn:Transport>
\end{lstlisting}

\subsection{VOEvent message receipt response}
\label{sec:transport:ack}

The VOEvent message receipt response is similar to the \xmlel{iamalive}
response except the role is either \xmlel{ack} or \xmlel{nak}, the
\xmlel{<Origin~/>} is the IVOID of the just-received VOEvent message, and an
optional \xmlel{<Result~/>} element may accompany the \xmlel{<Param~/>}
elements. \xmlel{<Result~/>} may contain any string; it is recommended that it
contain a human-readable error message if role is \xmlel{nak}. The
\xmlel{<TimeStamp~/>} element contains the date and time at which the response
was generated, formatted as per \S3.3.7 of \citet{Peterson:2012}. This time
should be provided in UTC, and may include a “Z” timezone indicator.

The \xmlel{nak} response indicates that the recipient is unable or unwilling
to take responsibility for this message. This may be because, for example, the
message fails to validate as a valid VOEvent, or because it was received from
an unauthorized client (\S\ref{sec:limit}). A \xmlel{nak} response is not
appropriate if the sender is able to accept the message but then decides not
to redistribute it (for example, if it is a duplicate of an event which has
already been distributed: Section \ref{sec:dedup}).

A \xmlel{nak} response should be regarded as a permanent failure state:
delivery of the VOEvent message which was met with the \xmlel{nak} to the
recipient which replied with the \xmlel{nak} should be aborted. If the
VOEvent message was being transmitted over an author-to-broker connection,
the author may identify the cause of the failure (for example by making use of
the information in the \xmlel{<Meta />} element of the \xmlel{nak}),
construct a corrected VOEvent message and attempt a repeat submission. If the
VOEvent message was being transmitted over a broker-to-subscriber connection,
the broker should abandon the attempt to deliver this message.

\begin{lstlisting}[language=XML,label=lst:ack,
                   caption=Sample VOEvent message receipt response indicating successful transmission (\xmlel{ack}).]
<?xml version='1.0' encoding='UTF-8'?>

<trn:Transport role="ack" version="1.0"
 xmlns:trn="http://telescope-networks.org/schema/Transport/v1.1"
 xmlns:xsi="http://www.w3.org/2001/XMLSchema-instance"
 xsi:schemaLocation="http://ivoa.net/xml/Transport/v1.1
                     http://ivoa.net/xml/Transport-v1.1.xsd">
    <Origin>ivo://invalid.author/example#0123456789</Origin>
    <Response>ivo://invalid.subscriber/example#</Response>
    <TimeStamp>2001-01-01T00:00:00Z</TimeStamp>
    <Meta>
        <Param name="IPAddr" value="10.0.0.0" />
        <Param name="Contact" value="name@subscriber.invalid" />
        <Result>Message received and validated successfully</Result>
    </Meta>
</trn:Transport>
\end{lstlisting}

\begin{lstlisting}[language=XML,label=lst:nak,
                   caption=Sample VOEvent message receipt response indicating unsuccessful transmission (\xmlel{nak}).]
<?xml version='1.0' encoding='UTF-8'?>

<trn:Transport role="nak" version="1.0"
 xmlns:trn="http://telescope-networks.org/schema/Transport/v1.1"
 xmlns:xsi="http://www.w3.org/2001/XMLSchema-instance"
 xsi:schemaLocation="http://ivoa.net/xml/Transport/v1.1
                     http://ivoa.net/xml/Transport-v1.1.xsd">
    <Origin>ivo://invalid.author/example#0123456789</Origin>
    <Response>ivo://invalid.subscriber/example#</Response>
    <TimeStamp>2001-01-01T00:00:00Z</TimeStamp>
    <Meta>
        <Param name="IPAddr" value="10.0.0.0" />
        <Param name="Contact" value="name@subscriber.invalid" />
        <Result>Error in VOEvent message: ISOTime not in ISO 8601 format</Result>
    </Meta>
</trn:Transport>
\end{lstlisting}

\section{Protocol operation}
\label{sec:protocol}

This section describes the operation and sequencing of VTP operation for each
end of a connection between an author and a broker, as well as between a
broker and a subscriber. See Section \ref{sec:node} above for a qualitative
discussion of the protocol from the viewpoint of each entity.

\subsection{Author sending to broker}
\label{sec:protocol:authortobroker}

\begin{figure}
  \begin{center}
  \includegraphics{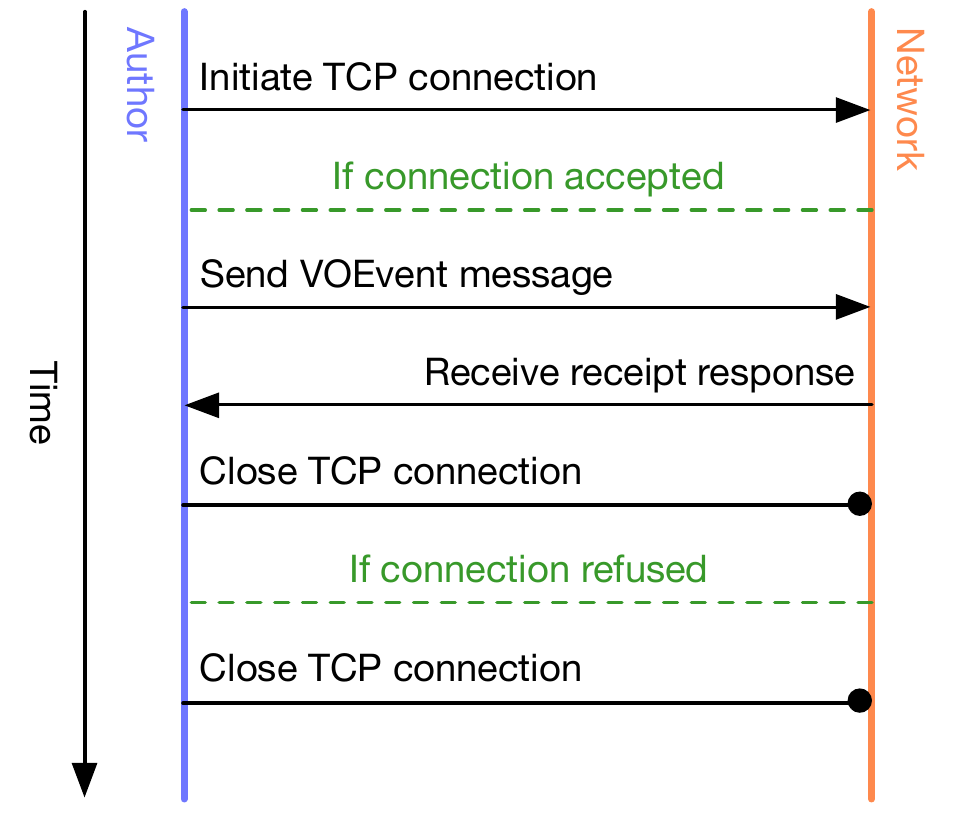}
  \end{center}

  \caption{Transport protocol at an author node.}

  \label{fig:protocol:authortobroker}
\end{figure}

The author initiates a TCP connection to the broker. The broker may choose to
accept or reject that connection based, for example, on an access control
whitelist (\S\ref{sec:limit:whitelist}). If the author is rejected, the
connection is terminated. If the connection is accepted, the author creates a
VOEvent message by prepending a byte count to the VOEvent document
(\S\ref{sec:common:format}) and transmits it to the broker. The author should
then wait for a VOEvent message receipt response (\S\ref{sec:transport:ack})
from the broker. The author may use this to determine whether the message
has been successfully delivered. The connection is then closed.

If a receipt response is not received, the author should assume that a
temporary failure has prevented the broker from accepting the message for
distribution. The author may close the connection and retry.

This transaction is illustrated in Figure \ref{fig:protocol:authortobroker}.

\subsection{Broker receiving from author}
\label{sec:protocol:brokerfromauthor}

\begin{figure}
  \begin{center}
  \includegraphics{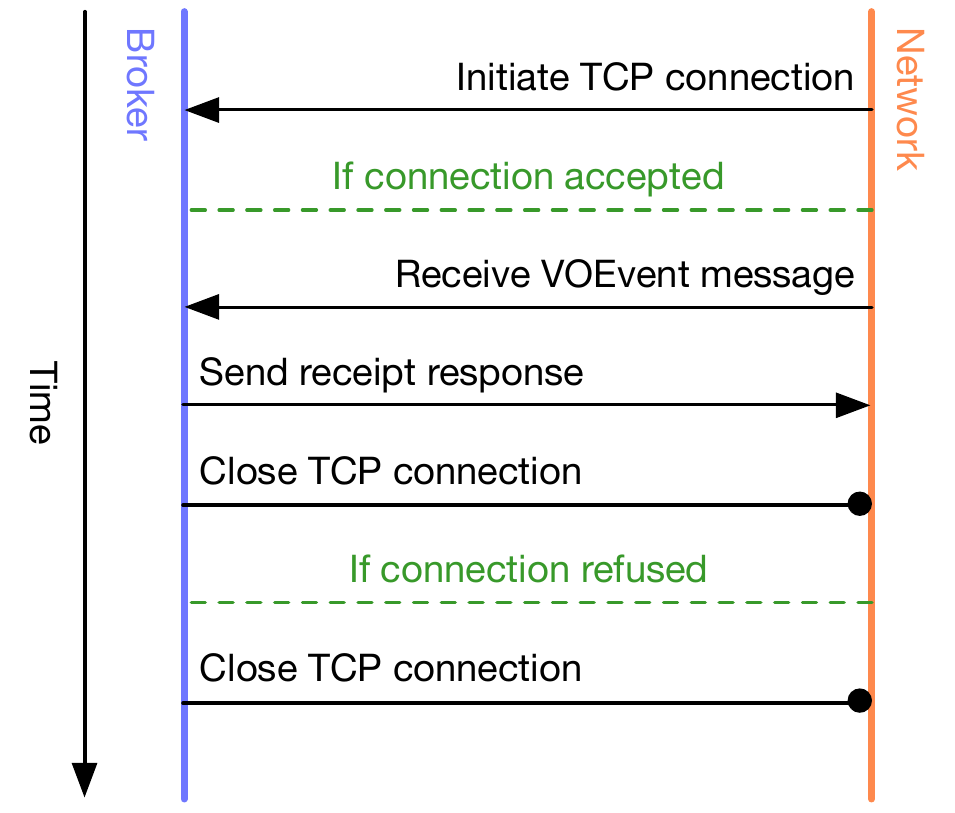}
  \end{center}

  \caption{Transport protocol at broker receiving from author.}

  \label{fig:protocol:brokerfromauthor}
\end{figure}

The broker awaits incoming TCP connections from authors. When a connection is
received, the broker may choose to accept or reject the connection based, for
example, on an access control whitelist (\S\ref{sec:limit:whitelist}). If the
author is rejected, the connection is terminated. Otherwise, the broker waits
to receive a VOEvent message from the author. When the VOEvent is received,
the broker should test the message for validity. The broker must return a
VOEvent message receipt response (\S\ref{sec:transport:ack}) to the author
indicating that it has either accepted (\xmlel{ack}) or refused (\xmlel{nak})
the VOEvent message. The connection is then closed.

This transaction is illustrated in Figure \ref{fig:protocol:brokerfromauthor}.

\subsection{Broker sending to subscriber}
\label{sec:protocol:brokertosub}

\begin{figure}
  \begin{center}
  \includegraphics{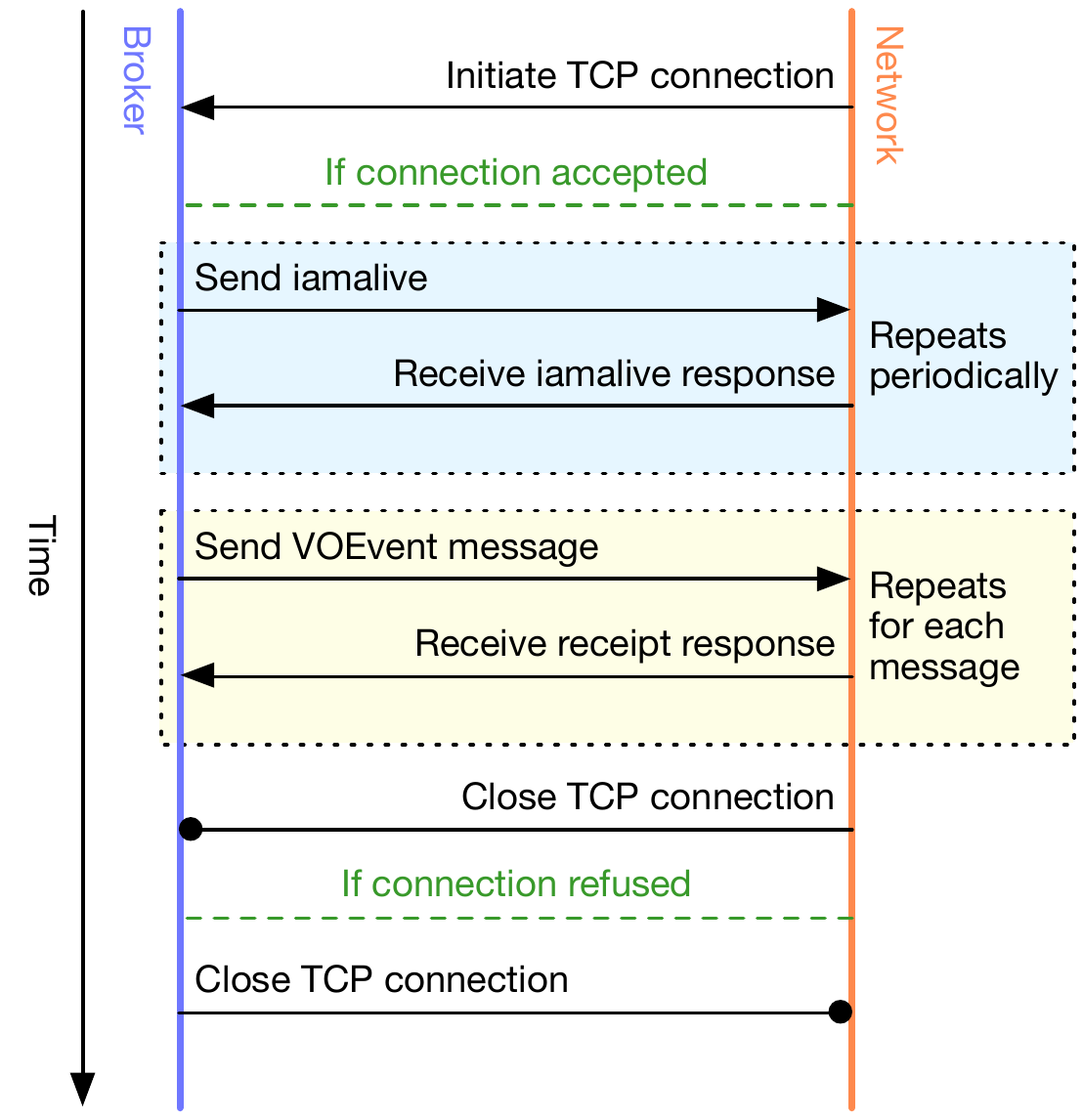}
  \end{center}

  \caption{Transport protocol at broker sending to subscriber.}

  \label{fig:protocol:brokertosub}
\end{figure}

The broker awaits incoming TCP connections from subscribers. When a new
connection is received, the broker may choose to accept or reject the
connection based, for example, on an access control whitelist
(\S\ref{sec:limit:whitelist}). If the subscriber is rejected, the connection
is terminated. Otherwise, the broker adds the subscriber to its distribution
list.

Periodically, at intervals of no more than 90\,s (\S\ref{sec:maintenance}),
the broker must send an \xmlel{iamalive} message
(\S\ref{sec:transport:iamalive}) to the subscriber. The subscriber must reply
with an \xmlel{iamalive} response (\S\ref{sec:transport:iamaliveresponse}).
If the broker does not receive \xmlel{iamalive} response messages from the
subscriber in a timely fashion, it may assume that the subscriber is dead or
gone and close TCP connection.

When the broker has a new VOEvent message ready for distribution, it is sent
to the subscriber. The broker receives a VOEvent message receipt response
(\S\ref{sec:transport:ack}) in reply. The broker may use this to determine
whether the VOEvent message has been accepted (\xmlel{ack}) or refused
(\xmlel{nak}). The broker must not attempt to repeat delivery of the message
if a \xmlel{nak} is received.

If a receipt response is not received, the broker should assume that a
temporary failure has prevented the subscriber from accepting the message. The
broker may attempt redelivery. At the discretion of the broker, repeated
failures to receive timely receipt responses may be grounds to terminate the
connection.

The TCP connection remains open, and the \xmlel{iamalive} exchange
continues, until either the subscriber explicitly closes the connection or
stops sending \xmlel{iamalive} response messages, thereby implicitly
indicating that the connection is closed.

These transactions are illustrated in Figure \ref{fig:protocol:brokertosub}.

\subsection{Subscriber receiving from broker}
\label{sec:protocol:subfrombroker}

\begin{figure}
  \begin{center}
  \includegraphics{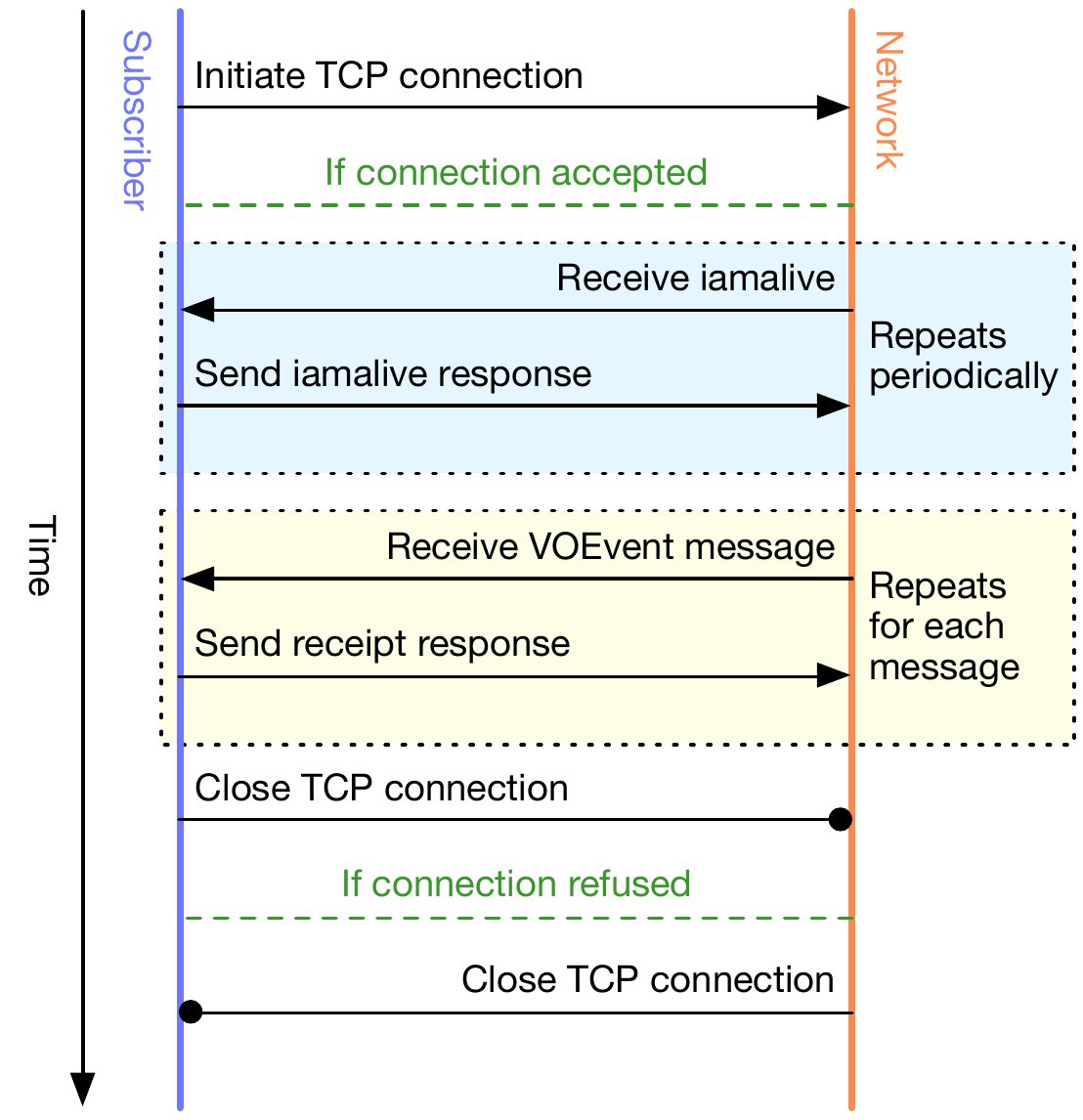}
  \end{center}

  \caption{Transport protocol at subscriber.}

  \label{fig:protocol:subfrombroker}
\end{figure}

The initiates a TCP connection to the broker. The broker may choose to accept
or reject that connection based, for example, on an access control whitelist
(\S\ref{sec:limit:whitelist}). If the subscriber is rejected, the connection
is terminated. Otherwise, the connection remains open, and the subscriber
begins receiving messages.

Periodically, at intervals of no more than 90\,s (\S\ref{sec:maintenance}),
the subscriber should expect to receive an  \xmlel{iamalive} message
(\S\ref{sec:transport:iamalive}) from the broker. The subscriber must reply
with an \xmlel{iamalive} response (\S\ref{sec:transport:iamaliveresponse}).
If the subscriber does not receive \xmlel{iamalive} messages from the broker
in a timely fashion, it may assume that the broker is dead or gone and close
TCP connection. The subscriber may attempt to re-connect to the broker.
Re-connection attempts should use a geometric back-off algorithm.

When a VOEvent message is received, the subscriber may test it for
validity. The subscriber must return a VOEvent message receipt response
(\S\ref{sec:transport:ack}) to the broker indicating that it has either
accepted (\xmlel{ack}) or refused (\xmlel{nak}) the VOEvent message.

The TCP connection remains open, and the \xmlel{iamalive} exchange
continues, until either the subscriber explicitly closes the connection or
stops receiving \xmlel{iamalive} messages, which indicates that the
connection has been terminated.

These transactions are illustrated in Figure \ref{fig:protocol:subfrombroker}.

\section{De-duplication}
\label{sec:dedup}

In a network topology like that illustrated in Figure \ref{fig:network},
multiple brokers service potentially overlapping sets of authors and
subscribers. As per Sections \ref{sec:intro} and
\ref{sec:node:brokertobroker}, brokers will subscribe to each other's event
feeds to ensure that their subscribers have access to the full range of
available events.

In this situation, there is a risk of event loops developing on the network:
broker A receives an event from B and forwards it to its subscriber list,
which includes A, which forwards it to its subscriber lists, which includes B,
and so on. In order to prevent this, each broker must process each unique
VOEvent message it receives a maximum of once.

Note that it is now established practice to distribute different descriptions
(e.g. VOEvent 1.1 and 2.0) of the same celestial event with the same IVOID
\footnote{Refer to
\url{http://www.ivoa.net/pipermail/voevent/2012-March/002836.html} and
subsequent discussion. Note that the VOEvent standard uses the now-deprecated
term “IVORN” in place of IVOID.}. Consequently, an IVOID is not a unique
identifier of a particular VOEvent message, and is not, therefore, suitable
for use in network de-duplication.

Instead, we regard two messages as being the same if the content between the
opening \xmlel{<} and the closing \xmlel{>} of the \xmlel{<VOEvent~/>}
element is bit-for-bit identical, including all white space characters. The
implementation of this check is left to the discretion of the broker\footnote{
Appropriate techniques may include directly comparing the bitstream (which
would necessarily mean storing an archive of previously-processed events) or
calculating a hash function such as SHA1 \citep{Eastlake:2001} over the event
contents and storing the result.}.

In the event that some future revision of the VOEvent standard adopt an
identifier which is unique to the message, rather than to the celestial event,
it would be preferable to use that for de-duplication rather than calculating
a hash over the event content.

\section{Limiting access}
\label{sec:limit}

For administrative or security reasons, broker administrators may wish to
limit access to the services they provide to a restricted range of clients.
These restrictions may be required on either or both of the connection types
in VTP: author to broker (a limit on which authors can publish through a given
broker) or broker to subscriber (a limit on which subscribers a broker is
willing to provide with event streams).

Two mechanisms are may be applied within the VTP framework to address these
requirements.

\subsection{IP address whitelisting}
\label{sec:limit:whitelist}

If the broker knows a priori the IP addresses or ranges from which authorized
authors or subscribers are permitted to connect (a ``whitelist''), it may
simply deny connections from addresses which fall outside that range. If the
sets of authorized authors and subscribers are not the same, separate
whitelists may be implemented.

This mechanism imposes significant administrative overhead on the broker owner
if large and complex whitelists are required. Further, it is of limited
applicability if the clients to be serviced are using dynamic IP addresses
(that is, addresses which change periodically).

\subsection{Cryptographic signatures}
\label{sec:limit:crypto}

A digital signature scheme enables the recipient of a digital message to
verify the identity of its author \citep{Diffie:1976}. By requiring authors
and subscribers to apply appropriate signatures to VOEvent and Transport
messages, it may be possible for a broker to verify their identity and
restrict the services made available to them.

Various digital signature schemes which are appropriate for use with VOEvent
and other XML documents have been suggested \citep{Allen:2008, Denny:2008}. At
time of writing, none have seen significant adoption. Given that, this version
of the VTP standard does not specify a particular approach, nor require that
any form of cryptographic authentication be available. However, the
\xmlel{<Transport~/>} schema provides an \xmlel{<authenticate>} message type,
shown in Listing \ref{lst:authenticate}, which may be used to implement either
the \citeauthor{Allen:2008} or \citeauthor{Denny:2008} scheme, or as the basis
for some other approach.

See also \citet{std:SSOAUTH2} for a description of the authentication schemes
available for use across the Virtual Observatory.

\begin{lstlisting}[language=XML,caption=Sample \xmlel{authenticate} message.,
                   label=lst:authenticate]
<?xml version="1.0" encoding="UTF-8"?>

<trn:Transport role="authenticate" version="1.0"
 xmlns:trn="http://telescope-networks.org/schema/Transport/v1.1"
 xmlns:xsi="http://www.w3.org/2001/XMLSchema-instance"
 xsi:schemaLocation="http://ivoa.net/xml/Transport/v1.1
                     http://ivoa.net/xml/Transport-v1.1.xsd">
    <Origin>ivo://invalid.broker/example#</Origin>
    <TimeStamp>2001-01-01T00:00:00Z</TimeStamp>
</trn:Transport>
\end{lstlisting}

\newpage
\appendix

\section{Transport schema}
\label{sec:transportschema}

\begin{lstlisting}[language=XML]
<?xml version="1.0" encoding="utf-8" ?>
<xs:schema xmlns:xsi="http://www.w3.org/2001/XMLSchema-instance"
           xmlns:xs="http://www.w3.org/2001/XMLSchema">
 <xs:element name="Transport">
  <xs:complexType>
   <xs:sequence>
    <xs:element minOccurs="1" name="Origin" type="xs:anyURI" />
    <xs:element minOccurs="0" maxOccurs="1" name="Response" type="xs:anyURI" />
    <xs:element minOccurs="1" name="TimeStamp" type="xs:dateTime" />
    <xs:element minOccurs="0" maxOccurs="1" name="Meta">
     <xs:complexType>
      <xs:sequence minOccurs="1" maxOccurs="1">
       <xs:element minOccurs="0" maxOccurs="unbounded" name="Param">
        <xs:complexType>
         <xs:attribute name="name" type="xs:string" use="required" />
         <xs:attribute name="value" type="xs:string" use="required" />
        </xs:complexType>
       </xs:element>
       <xs:element minOccurs="0" maxOccurs="1" name="Result" type="xs:string" />
      </xs:sequence>
     </xs:complexType>
    </xs:element>
   </xs:sequence>
   <xs:attribute name="role" type="roleType" use="required" />
   <xs:attribute name="version" type="xs:string" use="required" />
  </xs:complexType>
 </xs:element>
 <xs:simpleType name="roleType">
  <xs:restriction base="xs:string">
   <xs:enumeration value="iamalive" />
   <xs:enumeration value="authenticate" />
   <xs:enumeration value="ack" />
   <xs:enumeration value="nak" />
  </xs:restriction>
 </xs:simpleType>
</xs:schema>
\end{lstlisting}

\section{Version history}
\label{sec:history}

\subsection{Revised since v2.0-PR-20161230}

\begin{itemize}

    \item{Split history by PR revision (Appendix \ref{sec:history}).}

    \item{Note that different implementations of VTP should be interoperable
    (\S\ref{sec:common:design}).}

\end{itemize}

\subsection{Revised since v2.0-PR-20160503}

\begin{itemize}

    \item{Replaced all references to ``IVORN'' with ``IVOID''.}

    \item{Make clear that authors and brokers should be registered with the
    IVOA registry (\S\ref{sec:node}).}

    \item{Substantially trimmed the material on cryptographic signatures
    (\S\ref{sec:limit:crypto}). Made it clear that this version of VTP does
    not specify a particular approach.}

    \item{Describe appropriate actions when a receipt response message is not
    received (\S\S\ref{sec:protocol:authortobroker} \&
    \ref{sec:protocol:brokertosub}).}

    \item{Indicate that timestamps should be in UTC
    (\S\S\ref{sec:transport:iamalive}, \ref{sec:transport:iamaliveresponse},
    \ref{sec:transport:ack}).}

\end{itemize}

\subsection{Revised since IVOA Note v1.1}

\begin{itemize}

    \item{Add Section \ref{sec:common:design}, describing design goals of the
    protocol.}

    \item{Add Section \ref{sec:dedup}, detailing requirements for message
    de-duplication to avoid network loops.}

    \item{Specify an explicit interval requirement to connection maintenance
    messages (\S\ref{sec:maintenance}).}

    \item{Clarify the semantics of \xmlel{nak} Transport messages
    (\S\ref{sec:transport:ack}).}

    \item{Make it explicit that brokers should not attempt to repeat delivery
    of messages which meet with a \xmlel{nak} on the first attempt: VTP does
    not support the concept of a ``temporary failure''
    (\S\S\ref{sec:transport:ack}, \ref{sec:protocol:brokertosub}).}

    \item{Reword the descriptions of protocol operation so that they describe
    only the traffic exchanged over the network and not the implementation of
    the various entities (\S\ref{sec:protocol}).}

    \item{Allow timezone specification in \xmlel{iamalive}
    \xmlel{<TimeStamp~/>} elements (\S\S\ref{sec:transport:iamalive},
    \ref{sec:transport:iamaliveresponse}, \ref{sec:transport:ack}).}

    \item{Remove identifying information from example XML documents
    (\S\S\ref{sec:transport:iamalive}, \ref{sec:transport:iamaliveresponse},
    \ref{sec:transport:ack}, \ref{sec:limit:crypto}).}

\end{itemize}

\subsection{Revised since IVOA Note v1.0}

\begin{itemize}

    \item{Add an optional \xmlel{<Result~/>} sub-element (containing text)
    within the optional \xmlel{<Meta~/>} element. This is intended to convey
    details on errors encountered if the Transport response is \xmlel{nak} but
    may also be used for informational purposes in \xmlel{ack} messages.}

\end{itemize}

\bibliography{ivoatex/ivoabib,vtp}

\end{document}